\documentclass[12pt,a4paper]{article}
\usepackage{graphicx}
\usepackage[latin1]{inputenc}
\usepackage{amsmath}
\usepackage{amsfonts}
\usepackage{amssymb}
\title{Glycol and Water Solubility. An Interacting Molecular Description. }
\author{Fredrick Michael*}

\begin{document}
\maketitle                                             
\begin{abstract}
 Recently there has been a growing interest in Glycol fluid solubility in water. The applications of Glycol-Water fluids is in the biomolecular field, physical chemistry, and engineering, from protein dynamics to power production and storage devices technology. In this paper a theoretical model for  nanometer scale or mesoscale interacting fluids is derived for the interacting Glycol and water molecules. The theory is a semi-classical  interacting potential theory which utilizes effective potentials such as of the Lennard-Jones type. The statistical distributions of the fluids are derived from information theory and temperature dependent distributions are obtained for the Lennard-Jones potential interactions of the fluids.

\end{abstract}

\section{Introduction}
 The applications of Glycol-Water fluids is in the biomolecular field, physical chemistry, and engineering, from protein dynamics to power production and storage devices technology. In this paper a theoretical model for  nanometer scale or mesoscale interacting fluids is derived for the interacting Glycol and water molecules. The theory is a semi-classical  interacting potential theory which utilizes effective potentials such as of the Lennard-Jones type. The statistical distributions of the fluids are derived from information theory and temperature dependent distributions are obtained for the Lennard-Jones potential interactions of the fluids.  

The full description of the fluids of water and Glycol would require a fully quantum mechanical theory. Such a theory exists and will be briefly discussed in the derivation. However for practical applications, simplifications of the computation required to model the biomolecular dynamics of Glycol-Water are sought. Simplifications are possible and are motivated by the success of Lennard-Jones molecular dynamics of fluids and biomolecular interactions, at nanometer scales. Such models the utilize Lennard-Jones type of potential interactions for example are incorporated in computational molecular dynamics such as the NAmd Nanomolecular dynamics source code free distributions of the University of Illinois molecular dynamics of biomolecules project \cite{namd1}.
%itlies,itspies,itdeludes,itattemptmurder,itistoremoved
%the imporatnt thing is to remove it

The theoretical model will be derived for the two fluids as independent fluids. Then the interacting binary fluids model will be developed and discussed. The resulting statistics will be numerically simulated and compared with experimental results published by researchers that have conducted detailed experimental studies of Glycol-Water interactions.

\section{Interaction between differing molecular species and phases}

The Glycol interacts with water by Hydrogen-Oxygen bonding. The Water is attracted to the Oxygen atoms and Hydrogen bonds are formed in the case of ethylene oxide adducts dissolved in water.
\begin{eqnarray}
\begin{matrix}  {-CH_2 -{O} -CH_2 -} \\ {|}\\{H}\\{|}\\{O}\\{|}\\{H}      \end{matrix}
\end{eqnarray}
A 1980 study by steuter, Mozafar and Goodin \cite{steuten1}  also has shown that some of the ethylene adducts in aqueous solutions 'may be present as cation-active polyoxonium compounds'
\begin{eqnarray}
\begin{matrix} \\{H}\\{|}\\  \;\;\;\;\;\;\;\;\;\;\;\;\; {{[-CH_2 -{O} -CH_2 -]^+} {(OH)^-}}       \end{matrix}
\end{eqnarray}
With these two types of reactions (1) and (2) co-existing in an equilibrium

%\pagebreak
\bigskip

\section{Water model}

Water is a fluid that is the universal solvent of nature . It is  a fluid at standard temperature and pressure. The fluidity of water is of a complex arrangement of dipole-dipole interactions of the  polar molecule. Water has a triangular shape with a bend angle of $\alpha = 104.5^{\small{\small{o}}}$. Water is comprised of $H-O-H$ connected by covalent bonds and forms Hydrogen bonds readily with most molecular material.

Water can be described by a Hamiltonian for fluids. The Hamiltonian is composed of the Hydrogen Hamiltonian and the Oxygen Hamiltonian and the interacting Hydrogen-Oxygen potential . The Hydrogen potential is $\hat H_{\text{H}} =  \hat h_{e^-} +   \hat h_{p^+} +\hat h_{e^-  e^-}+\hat h_{p^+  -p^+}  - \hat h_{e^-  p^+} $
\begin{eqnarray}
\hat{H_{\text{O}}}=  \sum \limits_{\vec k} ^{}  \epsilon_{\vec{k}} {c^\dagger}_{\vec{k}}{c}_{\vec{k}} +  \sum \limits_{\vec k'} ^{}  \tilde\epsilon_{\vec{k'}} {c^\dagger}_{\vec{k'}}{c}_{\vec{k'}}+
\sum \limits_{\vec k,\vec{l},\vec{m},\vec{n}}{} v({\vec{k},\vec{l},\vec{m},\vec{n}}) {c^\dagger}_{\vec{k}}{c}_{\vec{l}}{c^\dagger}_{\vec{m}}{c}_{\vec{n}} \\ \nonumber
+\sum \limits_{\vec k',\vec{l'},\vec{m'},\vec{n'}}{} \tilde v({\vec{k'},\vec{l'},\vec{m'},\vec{n'}}) {c^\dagger}_{\vec{k'}}{c}_{\vec{l'}}{c^\dagger}_{\vec{m'}}{c}_{\vec{n'}} -\sum \limits_{\vec k',\vec{k},\vec{l},\vec{l'}}{} \phi({\vec{k'},\vec{k},\vec{l},\vec{l'}}) {c^\dagger}_{\vec{k'}}{c}_{\vec{l}}{c^\dagger}_{\vec{k}}{c}_{\vec{l'}} \\  \nonumber
  .    \label{eq1} 
\end{eqnarray}
The gaseous atoms due to attraction form molecular bonds and become $H_2$. To illustrate the more complex interactions and the method of description we discuss the interaction formalism and the phase transition theory, which we will apply to the more complex $H_2 O$ in a similar approach, and to the Glycol molecular fluid and the Glycol-Water binary fluid. '

The Hamiltonian for Hydrogen can be written as a matrix
\begin{eqnarray}
\hat H_{\text{H}}=
 \left( \begin{matrix}{ \hat h_{e^-  e^-}} \:\:\:\:\: {\hat h_{e^-  p^+}}  \\ {\hat {h^\dagger}_{e^-  p^+}} \:\: \: {\hat h_{p^+  -p^+}}\end{matrix}\right)
\end{eqnarray}
where we have partitioned the Hamiltonian, a freedom we have in this and in all subsequent work, to include the free electron and electron-electron interaction as $\hat h_{e^- - e^-}$, similarly for the free proton and proton-proton interaction, and the interactions between the electron-proton as the off diagonal interaction terms. Note that the electron-proton interaction is of negative sign, a potential that lowers the energy to a more stable configuration, and also that the off diagonal terms are hermitian conjugates and transposes of each other.

 The Hydrogen molecule is then described by the long range order parameters, the diagonal DLRO parameter useful for observables such as the number of electrons and protons, as $N_{(e^-) , (p^+)} = N_{e^-} + N_{p^+}$, and the off-diagonal ODLRO useful for interacting formation of Hydrogen and Hydrogen Molecular gas as $N_{H + H_2} = N_{e^- - p^+}$. These are obtained from the retarded and advanced $r,a$ Green's functions of the Hamiltonians, which are for example for the electrons $g^{r,a}_{e^- - e^-}=\mp i \frac{f(\omega)}{\hbar \omega - \epsilon(\vec k) - i\sigma(\vec k \omega) \pm i\eta}$  and $\eta$ is an infinitesimal that shifts the singular denominator such that outgoing (retarded) wave solutions for example are obtained. Also, the Fermi-Dirac distribution $f(\omega)$ is a temperature dependent distribution that is a step function at zero temperature and is a limit on the energy integration $N=i\int g^r (\vec k, \omega) d\vec k  d\vec \omega$ by the dispersion relation. Other observables, such as energy $<E>$ and rates of reaction can also be obtained from this matrix formulation and the ODLRO as the derivative of the time dependent observables as  $\frac{d N_{H + H_2}}{dt} =\frac{\partial }{\partial t} i\int g^r (\vec k, t-t') d\vec k $. 

The phase transition of the molecular Hydrogen is described by another matrix interaction, the interaction being of the fluid and the lattice phases for example. Writing the Green's functions matrix as 
\begin{eqnarray}
\hat g_{lattice}=
\left( \begin{matrix}{ g_{f-f}} \:\: g_{f-S}  \:\: g_{f-g}\\ { g^{\dagger} _{S-f}} \:\: g_{S-S} \:\: g_{S-g} \\  g_{g-f} \:\: g_{g-S}  \:\: g_{g-g}\end{matrix}\right)
\end{eqnarray} 
The Hamiltonians are simplified Hydrogenic Hamiltonians $h_{g-g}=\sum\limits_{\vec k}^{} \epsilon(\vec k)  {c^\dagger}_{\vec{k}}{c}_{\vec{k}}+ \sum\limits_{\vec k,\vec l, \vec m, \vec n}^{} \phi(\vec k,\vec l, \vec m, \vec n) {c^\dagger}_{\vec{k}}{c}_{\vec{l}}{c^\dagger}_{\vec{m}}{c}_{\vec{n}} $, and the lattice Hamiltonian is shifted by  constant wave vectors corresponding to the lattice periodicity, geometry and symmetry as $h_{S-S}=\sum\limits_{\vec k + \vec Q_o}^{} \epsilon(\vec k)  {c^\dagger}_{\vec k+ \vec Q_o}{c}_{\vec{k}}+ \sum\limits_{\vec k + \vec Q_o,\vec l, \vec m - \vec {Q'}_o, \vec n}^{} \phi(\vec k,\vec l, \vec m, \vec n) {c^\dagger}_{\vec k+ \vec Q_o}{c}_{\vec{l}}{c^\dagger}_{\vec m- \vec {Q'}_o}{c}_{\vec{n}} $. The fluid phase Hamiltonian is similar as the lattice Hamiltonian however the constancy of the lattice vector is relaxed to $\vec Q_o -> \vec Q$ a variable. Other simplifications are possible, such as solid phase tight binding Hamiltonians, metallic effective mass Hamiltonians. 

The formation of water is a similar interaction description from this formulation of many-particle physics of interacting atoms and molecules and phase transitions .
The formation of the water molecule in the fluid phase is approached from the gaseous or vapor phase for definiteness. The Hydrogen gas of Hydrogen atoms and $H_2$ molecules is for example allowed into an Oxygen environment and the two gases interact as due to  Oxygen's effective electron affinity. The matrix Green's function for the formation of Water $H_2 O $ vapor then is written as 
\begin{eqnarray}
\hat g_{(H_2 O)_{vapor}}=
\left( \begin{matrix}{ g_{H-H}} \:\: g_{H-O}\\ { g^{\dagger} _{O-H}} \:\: g_{O-O} \end{matrix}\right)
\end{eqnarray}
and the Hamiltonians are written as discussed with the appropriate symmetry, approximation as needed to the required accuracy desired. 

The fluid phase is obtained as before , with the Hamiltonians including interaction terms of the phase symmetry, 
\begin{eqnarray}
\hat g_{(H_2 O)_{vapor-fluid}}=\left( \begin{matrix}  g_{v-v} \:\:  g_{v-f} \\  {g^{\dagger}}_{f-v} \:\: g_{f-f} \end{matrix}\right)
\end{eqnarray}

The Hamiltonians are of interactions between the vapor-fluid phases, and the interactions are of the gas potential with  the fluid potential vectors with the 'quasi-periodic' lattice vectors shifted by the variable vector $<\Phi>=\sum\limits_{\vec k + \vec Q ,\vec l, \vec k', \vec l'}^{} \phi(\vec k, \vec Q, \vec l, \vec k', \vec l') {c^{\dagger}}_{\vec k + \vec Q} c_{\vec k'}{c^{\dagger}}_{\vec l } c_{\vec l'}$ and here this denotes that a vapor 'particle' is created at $\vec l$ and propagated through the system , interacting with the fluid 'particle' of $\vec k + \vec Q$ the two particles then being removed at their final momentums of $\vec k' , \vec l'$ respectively and here we have written the fluid symmetry wave vector explicitly.

\section{Lennard-Jones theory of multi-phase fluids.}
The description of water from the fully quantum mechanical can be simplified. An indication of this was already given when we discussed the level of accuracy, approximations utilized to achieve this accuracy , such as the use of Hydrogenic wave functions, tight binding Hamiltonians, effective mass Hamiltonians... An approximation that can be made is an approximation of the dynamics as quantum effects become screened. As the number of atoms and molecules increases, the screening that occurs makes it less necessary to describe the interactions at the fully quantum mechanical description. The description of the interactions can be made then as semi-classical interactions. The most well-known of theses descriptions being the Lennard-Jones potential and its generalization $\phi(r)=        \frac{a}{r^\mu} - \frac{b}{r^{\nu}}$ description of interactions, which is a screened Coulomb potential with the Pauli exclusion principle accounted for by the infinite amount of energy needed for two particles to occupy the same space, this being a mathematical description of repulsion of a sort. The Lennard-Jones potential is further simplified by the approximation by a harmonic oscillator potential in the vicinity of the minima of the potential well. However this approximation we consider as not being of sufficient accuracy to describe the dynamics of the fluid phase of water, and the subsequent description of the dynamics of Glycol.

  Nanofluid droplets and the physics of fluids at the nanometer scale and at the meso scale between the fully quantum mechanical and the classical continuum of hydro-dynamics have been described reasonably successfully in the literature by semi classical methods \cite{semiclassical1,semiclassical2}. The  approaches have been to classically treat the nanofluid droplet by approximate hydrodynamic means, the vapor by Boltzmann equation, and to evolve the number of particles and obtain rates of change in number of particles. Approximations are made such as Boltzmann equation \cite{semiclassical2} collision terms of a certain order, an example is the force collision terms up to a third order expansion by \cite{russianauthor}. 
  
  In hydrodynamics approaches, full analytic solutions are not obtained to the hydrodynamic equations... approximations of linearization and similar assumptions for solution are made such as in \cite{semiclassical1} , examples of approximations are of viscosity and compressibility.

  Another approach between the hydrodynamic approximation and the quantum mechanical dynamics of the  mesoscale  is possible. Given the success of recent molecular dynamics simulations in describing experimental results, approximating the nanofluid as a Lennard-Jones fluid, this approach seems reasonable. Then a similar theory of multi-phase fluid-vapor can be made at the Lennard-Jones approximation level that is applicable in between the range of applicability of the Boltzmann and hydrodynamics equations' scale of applicability and the fully quantum mechanical dynamics. That is, from the hundreds of atoms to the micrometer scale. From an analogy to the fully quantum mechanical phase transition theory discussed in the previous section, a theory of coexisting phases of nanofluid droplets can be obtained. The advantages would be the simultaneous description of co-existing phases by analogy to the matrix formulation discussed, long range order parameter approach as in the diagonal trace and summation to obtain particle or number of atoms in the individual phases.

    A theoretical advantage of the full quantum theory aside from increased accuracy at the 'cost' of increased computation time is the generality of the theory and the well known Hamiltonian superposition principle that allows us to include external perturbations and interaction terms and which allows us to write a theory of phase transitions as a superposition of interactions between the co-existing phases. From the point of view of the approach of Lennard-Jones fluids coexisting phases theory, the superposition principle is replaced by the Legendre transform of the maximum entropy method of theory derivation, and equivalently the information theoretic method.

 Alternatively a Lennard-Jones semi classical Lagrangian or Hamiltonian may be used, and dynamics with or without a heat bath computed. The heat bath dynamics implying fluctuations and therefore stochastic dynamics, equivalent descriptions made by Fokker-Planck equations can be  obtained.  These are analogous to the quantum mechanical Green's functions Schroedinger-like equations and can therefore be written in matrix form for the co-existing phases. We derive our theory utilizing the entropy maximization method.

\subsection{Derivation}  
   The entropy state function or information measure to be used we choose to be the Gibbs-Boltzmann entropy, $<S>=-c \int{P(\vec{r},t)lnP(\vec{r},t)}{d\vec{r}}$ which can be generalized to the nonextensive entropy of C. Tsallis \cite{tsallis1,fred2} for better accuracy in future work....  also other novel effects such as the q-parametrization of the deviation of the PDF from the Gaussian to the power-law PDF can be advantageously made for simple interacting nonlinear dynamics. Note that nonlinearity due to  physical interactions deviates the PDF from the Gaussian distribution towards the power-law distribution. The nonextensive statistics has been noted for theoretically producing this form of power-law distribution from a formally similar mathematics to the traditional thermodynamics and information theory. The connections are more profound and review articles and applications can be found at \cite{tsallis1} .   
  
   The maximum entropy approach is then of the maximization of the entropy measure constrained with observables of the system in a Legendre transform with Lagrange multipliers setting the weighted units to the entropic measure's.

   The choice of observables is one of choosing the minimum set of observables that describe the physics of interest. These observables are  described mathematically as functions of the other observables chosen as representative of the observable dynamics of the system. A discussion of this was made by the original author of maximum entropy theory \cite{jaynes1, jaynes2} E. T. Jaynes.  
 
  We choose observables analogous to the quantum mechanical observables in the fully quantum mechanical theory. The momentum or kinetic energy observables of quantum mechanics correspond to the first and second moments of the variables of position, the potential energy or interaction potential terms are similarly included here though we choose the Lennard-Jones potential and not the Coulomb potential,
  The observable moments are:
\begin{eqnarray}  
 {_H}M^2 (\Delta\vec x_{ij})= \sum \limits_{i,j}^{N_H}    <(\Delta \vec x_{ij} - <\Delta\vec x_{ij}>)^2> \nonumber \\ = \sum \limits_{i,j}^{N_H}\int (\Delta\vec x_{ij} - <\Delta\vec x_{ij}>)^2 P(\Delta\vec{x_{ij}},\Delta\vec{x_{i'j'}},......,\Delta{t}) {d\Delta\vec{x_{ij}}}    
 \end{eqnarray}

\begin{eqnarray}  
 {_O}M^2 (\Delta\vec x_{i'j'})= \sum \limits_{i',j'}^{N_O}    <(\Delta \vec x_{i'j'} - <\Delta\vec x_{i'j'}>)^2> \nonumber \\ = \sum \limits_{i',j'}^{N_O}\int (\Delta\vec x_{i'j'} - <\Delta\vec x_{i'j'}>)^2 P(\Delta\vec x_{ij},\Delta\vec{x_{i'j'}},......,\Delta{t}) {d\Delta\vec{x_{i'j'}}}    
  \end{eqnarray}

\begin{eqnarray}  
 {_OH}M^2 (\Delta\vec x_{k''k'})= \sum \limits_{k'',k'}^{N_OH}    <(\Delta \vec x_{k''k'} - <\Delta\vec x_{k''k'}>)^2> \nonumber \\ = \sum \limits_{k'',k'}^{N_OH}\int (\Delta\vec x_{k''k'} - <\Delta\vec x_{k''k'}>)^2 P(\Delta\vec x_{ij},\Delta\vec{x_{i'j'}},......,\Delta{t}) {d\Delta\vec{x_{i'j'}}}    
  \end{eqnarray}

 \begin{eqnarray}
 <V_{H-H}>= \sum \limits_{i,j}^{N_H}  < ( { \frac{\sigma_{H-H}} {{r_{ij}}^{\mu}} }  - { \frac{\eta_{H-H}} { r_{ij}^{\nu}} } )> \nonumber \\
 = \sum \limits_{i,j}^{N_H}  \int ( { \frac{\sigma_{H-H}} {{r_{ij}}^{\mu}} }  - { \frac{\eta_{H-H}} { r_{ij}^{\nu}} } ) P(\Delta\vec x_{ij},\Delta\vec{x}_{i'j'},.....,\Delta t) {d\Delta\vec{ x}_{ij}}{ d\Delta\vec{x}_{i'j'}}\end{eqnarray}

 \begin{eqnarray}
 <V_{O-O}>= \sum \limits_{i,j}^{N_O}  < ( { \frac{\sigma_{O-O}} {{r_{ij}}^{\mu}} }  - { \frac{\eta_{O-O}} { r_{ij}^{\nu}} } )> \nonumber \\
 = \sum \limits_{i,j}^{N_O}  \int ( { \frac{\sigma_{O-O}} {{r_{ij}}^{\mu}} }  - { \frac{\eta_{O-O}} { r_{ij}^{\nu}} } ) P(\Delta\vec x_{ij},\Delta\vec{x}_{i'j'},......,\Delta t) {d\Delta\vec{ x}_{ij}}{ d\Delta\vec{x}_{i'j'}} \end{eqnarray}

 \begin{eqnarray}
 <V_{H-O}>= \sum \limits_{i,j'}^{N_H,N_O}  < ( { \frac{\sigma_{H-O}} {{r_{ij'}}^{\mu}} }  - { \frac{\eta_{H-O}} { r_{ij'}^{\nu}} } )> \nonumber \\
 = \sum \limits_{i,j'}^{N_H,N_O}  \int ( { \frac{\sigma_{H-O}} {{r_{ij'}}^{\mu}} }  - { \frac{\eta_{H-O}} { r_{ij'}^{\nu}} } ) P(\Delta\vec x_{ij},\Delta\vec{x}_{i'j'},......,\Delta t) {d\Delta\vec{ x}_{ij}}{ d\Delta\vec{x}_{i'j'}} \end{eqnarray}

 \begin{eqnarray}
 <V_{{OH}-O}>= \sum \limits_{k',j'}^{N_{OH},N_O}  < ( { \frac{\sigma_{{OH}-O}} {{r_{k'j'}}^{\mu}} }  - { \frac{\eta_{{OH}-O}} { r_{k'j'}^{\nu}} } )> \nonumber 
\end{eqnarray}

 \begin{eqnarray}
 <V_{{H2O}-H2O}>= \sum \limits_{k,l}^{N_{H2O}}  < ( { \frac{\sigma_{{H2O}-H2O}} {{r_{kl}}^{\mu}} }  - { \frac{\eta_{{H2O}-H2O}} { r_{kl}^{\nu}} } )> \nonumber 
\end{eqnarray}

 \begin{eqnarray}
 <V_{{H2O}-Y}>= \sum \limits_{k,l'}^{N_{H2O},N_Y}  < ( { \frac{\sigma_{{H2O}-Y}} {{r_{kl'}}^{\mu}} }  - { \frac{\eta_{{H2O}-Y}} { r_{kl'}^{\nu}} } )> \nonumber 
\end{eqnarray}

 \begin{eqnarray}
 <V_{{H}-{( [Y]^+ (OH)^- )} }>= \sum \limits_{i,s}^{N_{H},N_{ ([Y]^+ (OH)^- )}}  < ( { \frac{\sigma_{{H}-{( [Y]^+ (OH)^- )}}} {{r_{is}}^{\mu}} }  - { \frac{\eta_{{H}-{( [Y]^+ (OH)^- )}}} { r_{is}^{\nu}} } )> \nonumber 
\end{eqnarray}

These observables of free 'particle' movement of trajectories and thermal fluctuations  and electromagnetic interactions as described by the semi-classical potentials of a Lennard-Jones type whose semi-classical approximation is general enough to have force constants that can be determined from experiment or alternative theoretical derivation. The interaction between particles of atoms forms water, and ions of Hydroxide and molecules of $H_2$ and $O_2$ and interacting $H_2 O$ molecules and interactions of water molecules with the Glycol's ether Oxygen atoms here denoted by species $Y$.

The maximization of entropy is made with the Lagrange multipliers setting appropriate units for the observables as $\delta[<S>]+\delta[\beta(a_H {_H}M^2 + .....+ a_{{H}-{( [Y]^+ (OH)^- )} } <V_{{H}-{( [Y]^+ (OH)^- )} }>]==0$ which derives the least biased distribution matrix for binary interactions
\begin{eqnarray}
\left ( {   \begin{matrix}  { {P_{H-H}}\;\;\;\;\;\;\;\;\;\;\;\;\;\;\;\; {....} \;\; {....} \;\;\;\;\;\;\;\;\; {P_{{H}-{( [Y]^+ (OH)^- )} } }}\\  {.....} \\{....}\\{....}\\{ {P_{{H}-{( [Y]^+ (OH)^- )} }} \;\;{....} \;\; {....}\;\; {...} \;\;{P_{{( [Y]^+ (OH)^- )} -{( [Y]^+ (OH)^- )} }}  } \end{matrix}} \right )
  \end{eqnarray}

We will specialize this to the simple case of fluid water solvent and dissolved ether, the steps in derivation thus far being instructive as how to proceed.

\section{ Lennard-Jones Water-Glycol interactions}
The simplified case of the water and dissolved adducts of the Glycol that is the ether and ethylene can be written as the previous derivation which we omit. We write the least biased distributions
\begin{eqnarray}
P(\Delta \vec{x}_{ij}, .....,\Delta t)=A_{(H_2 O ,  H_2 O)} \;\; e^{-\beta{ [ (\Delta \vec{x}_{ij}- <\Delta \vec{x}_{ij}>)^2 + a_{(H_2 O , H_2 O)}  ( { \frac{\sigma_{{H2O}, H2O}} {{r_{ij}}^{\mu}} }  - { \frac{\eta_{{H2O}, H2O}} { r_{ij}^{\nu}} } ) }]}  \:\:\:\:\:\:\:\: \:\:\:\:\:\:\:\:  \:\:\:\:\:\:\:\:  \\  \nonumber
.   \:\:\:\:\:\:\:\:     \:\:\:\:\:\:\:\: \:\:\:\:\:\:\:\: \:\:\:\:\:\:\:\:  \:\:\:\:\:\:\:\: \:\:\:\:\:\:\:\: \:\:\:\:\:\:\:\: \:\:\:\:\:\:\:\:  \:\:\:\:\:\:\:\: \:\:\:\:\:\:\:\: \:\:\:\:\:\:\:\: \:\:\:\:\:\:\:\:\\ \nonumber
.   \:\:\:\:\:\:\:\:     \:\:\:\:\:\:\:\: \:\:\:\:\:\:\:\: \:\:\:\:\:\:\:\:  \:\:\:\:\:\:\:\: \:\:\:\:\:\:\:\: \:\:\:\:\:\:\:\: \:\:\:\:\:\:\:\:  \:\:\:\:\:\:\:\: \:\:\:\:\:\:\:\: \:\:\:\:\:\:\:\: \:\:\:\:\:\:\:\:\\ \nonumber
.   \:\:\:\:\:\:\:\:     \:\:\:\:\:\:\:\: \:\:\:\:\:\:\:\: \:\:\:\:\:\:\:\:  \:\:\:\:\:\:\:\: \:\:\:\:\:\:\:\: \:\:\:\:\:\:\:\: \:\:\:\:\:\:\:\:  \:\:\:\:\:\:\:\: \:\:\:\:\:\:\:\: \:\:\:\:\:\:\:\: \:\:\:\:\:\:\:\:\\ \nonumber
.   \:\:\:\:\:\:\:\:     \:\:\:\:\:\:\:\: \:\:\:\:\:\:\:\: \:\:\:\:\:\:\:\:  \:\:\:\:\:\:\:\: \:\:\:\:\:\:\:\: \:\:\:\:\:\:\:\: \:\:\:\:\:\:\:\:  \:\:\:\:\:\:\:\: \:\:\:\:\:\:\:\: \:\:\:\:\:\:\:\: \:\:\:\:\:\:\:\:\\ \nonumber
.   \:\:\:\:\:\:\:\:     \:\:\:\:\:\:\:\: \:\:\:\:\:\:\:\: \:\:\:\:\:\:\:\:  \:\:\:\:\:\:\:\: \:\:\:\:\:\:\:\: \:\:\:\:\:\:\:\: \:\:\:\:\:\:\:\:  \:\:\:\:\:\:\:\: \:\:\:\:\:\:\:\: \:\:\:\:\:\:\:\: \:\:\:\:\:\:\:\:\\ \nonumber
P(\Delta \vec{x}_{i'j'}, ..... , \Delta t)=  \:\:\:\:\:\:\:\:     \:\:\:\:\:\:\:\: \:\:\:\:\:\:\:\: \:\:\:\:\:\:\:\:  \:\:\:\:\:\:\:\: \:\:\:\:\:\:\:\: \:\:\:\:\:\:\:\: \:\:\:\:\:\:\:\:  \:\:\:\:\:\:\:\: \:\:\:\:\:\:\:\: \:\:\:\:\:\:\:\: \:\:\:\:\:\:\:\:  \:\:\:\:\:\:\:\: \:\:\:\:\:\:\:\:  \:\:\:\:\:\:\:\: \:\:\:\:\:\:\:\:  \:\:\:\:\:\\ \nonumber
 A_{(H ,{( [Y]^+ (OH)^- )}  )} \;\; e^{-\beta{ [ (\Delta \vec{x}_{i'j'}- <\Delta \vec{x}_{i'j'}>)^2 ]} + a_{( H,[Y]^+ (OH)^- )}  ( { \frac{\sigma_{{{H},{( [Y]^+ (OH)^- )} }}} {{r_{i'j'}}^{\mu}} }  - { \frac{\eta_{{{H},{( [Y]^+ (OH)^- )} }}} { r_{i'j'}^{\nu}} } )  } \:\:\:\:\:\:\:\:   \:\:\:\:\:  \\  \nonumber
\end{eqnarray}
 The Lagrange multipliers are $\beta =\frac{1}{2D\Delta t}$ which are the inverse of the diffusion coefficient and the Diffusion constant is by the Einstein relation proportional to the inverse temperature from thermodynamics. The normalization is the inverse of the partition function and can be evaluated in several methods, we will leave that for later as it can be normalized arbitrarily or to unity.
   The solution of this least biased distribution is dependent upon the Lagrange multipliers and if normalization is needed then integration over the variables to obtain the partition function. Some identities utilized for evaluation are $A(t)=\frac{1}{Z(t)}$ the normalization and partition functions , $\beta(t) Z(t)^2 = const.$ and $\frac{1}{Z}\frac{\partial Z}{\partial \beta}=-< (\Delta \vec{x}_{i'j'}- <\Delta \vec{x}_{i'j'}>)^2>$ and similarly for other Lagrange multipliers and observables.

\pagebreak

\subsection{Derivation of Solutions}

\begin{equation}
  P(\Delta\vec x_{ij},\Delta t)= \frac{ e^{-\beta (\Delta\vec x_{ij} - <\Delta\vec x_{ij}>)^2  - a ( \frac{\sigma_{vv}} {{r_{ij}}^{\mu}}  - \frac{\eta_{vv}}{ r_{ij}^{\nu}} ) } } { Z(\Delta t)} . 
\end{equation}      

   The Lagrange multipliers are evaluated as discussed from the relationship between partition function and the observables $\frac{\partial lnZ}{\partial \beta}=-M^2 (\Delta\vec x_{ij})$ and similarly for the other observables. Note that the Lagrange multipliers of the second moments are related to the diffusion coefficients by the temperature-diffusion coefficient thermodynamics relation $D=\frac{1}{2\beta}$, and $<(x_{ij})^2)>=2Dt$. 
   The analytic solution of interest here is the one of the transformation of the Lennard-Jones potential to a drift coefficient term. This is done by uncompleting the square of the Gaussian
   \begin{eqnarray}
    p(X)= e^{  (X^2 - 2 X<X> + <X>^2)/2Dt}
    \end{eqnarray}
  and comparing it with the $r_{ij}$ dependent terms of the de-mean (or constant or linear and renormalized mean) Gaussian

     \begin{equation} 
    P(r_{ij},....)= e^{-\beta(  (r_{ij}^2 - 2 r_{ij} <r_{ij}> + <r_{ij}>^2 )- \gamma LJ(r_{ij})  )   }e^{.....}e^{.....}.....
   \end{equation}
   The Lennard-Jones potential terms can be seen to be terms that can be transformed to a drift coefficient or the first moment by un-completing the square. We obtain 
   \begin{equation}
   a(\vec r_{ij})=< r_{ij}>={r_{ij}}  + \sqrt{ {r_{ij} }^2    +  \frac{\gamma}{\beta} LJ(r_{ij}) }.  \label{LJdrift1}
   \end{equation}
   
 \subsection{Analytic distribution functions}  
   With this the $r_{ij}$ dependent terms become a simple $e^{-\beta (r_{ij}- a(r_{ij},\Delta t))^2 }$ Gaussian which by the identity  $\frac{\partial lnZ}{\partial \beta}=-M^2 (\Delta\vec x_{ij})$ gives the simple diffusion-temperature relation $\beta=\frac {1} {2 D_{r} t}$ with the temperature from thermodynamics proportional to the diffusion constant.
   
     The resulting PDF is a Gaussian with the coordinates expressed in spherical coordinates for simplicity. The integration for the partition function or normalization can be accomplished, however we use the relationship $\beta Z(t)^2=2\pi $ and the resulting distribution is
   \begin{equation}
  P(r_{ij},....)=  A_r (t)\frac {     e^{  \frac{-  (\vec r_{ij} -a(\vec r_{ij}) )^2}  {2{D}t} }  
   }  {  \sqrt{4\pi {D} t} ^2}.
  \end{equation} 
where the derivation implies 

a) the Lagrange multiplier second moment are related as $<r^2>=2Dt=\frac{1}{\beta}$

b) the inverse partition function and the normalization are therefore also $ Z_r (t) =  {  \sqrt{4\pi {D} t} ^2}$

c) the partition function $ Z(t)=z_r(t) z_{\theta} (t) z_{\phi} (t)$ can be factorized and,

d) Lagrange multiplier and factorized partition functions are also constants as the overall identity $\beta Z^2 = constant$.

 Note that aside from the $r_{ij}$ terms which have the complicated Lennard-Jones potential dependence and its radial coordinates dependence Eq.(\ref{LJdrift1}) added to the radial coordinate as a drift coefficient here a function, the other variables such as the velocity magnitudes in different coordinates are all simply zero mean $<\theta_{ij}>=0$, constant drift $<\theta_{ij}>=b$ or linear drift terms corresponding to accelerated motion $<\theta_{ij}>= c\theta_{ij}+b$; Note that a constant drift value mean is useful in uniform motion descriptions, and a linear $<\theta_{ij}>= c\theta_{ij}+b$  drift term corresponds to a constant driving force term. 

\section{Stochastic Derivation }
The derivation of the PDFs of interacting molecules can be made from the stochastics...the deterministic dynamics are the force and potential here the Lennard-Jones -like potential. The force after variation via the Euler-Lagrange equation is for the radial coordinate and here we mention that the different species of molecules are accounted for by the center of mass and coordinate separation 
$r= x_1 - x_2$ and $M= \frac{m_1 m_2}{m_1 + m_2}$ and
\begin{equation}
M \frac{dv_r}{dt}= -\frac{\partial LJ(r)}{\partial r}
\end{equation}
which can be solved for the radial coordinate and fluctuations related to the temperature and thermodynamics added by the diffusion-temperature Einstein relation, and since $ v=\frac{dr}{dt}$ 

\begin{equation}
dr= a(r,t)dt + \sqrt{D}dW(t)  \label{eqnstoch1}
\end{equation}

and here the velocity has been integrated and the drift coefficient is the integral w.r.t. time of the differential of the Lennard-Jones   -like potential  $ a(r,t)= \int \limits_{0}^{t} -\frac{\partial LJ(r)}{M \partial r} dt' $  and the fluctuations are a Wiener process' differential , and is delta correlated as a Gaussian white noise for simplicity $ <W(t)W(t')>=\delta (t-t') $...future research can generalize the fluctuations to the nonextensive statistics for very good Xi-squared experimental data fitting and therefore accuracy.

The stochastic differential equation Eq.(\ref{eqnstoch1}) is equivalent to the Fokker-Planck equation 

\begin{equation}
\frac{\partial p}{\partial t}=-\frac{\partial }{\partial r}[ a(r,t) p] + \frac{D}{2}\frac{\partial^2 p}{\partial r^2}
\end{equation}
 this equation is solved by the short time transition probability
\begin{equation}
p(r_{ij},r'_{ij},t,t')=\frac{e^{-\frac{(r_{ij} - r'_{ij} - a(r'_{ij},t')(t-t'))}{2D(t-t')}}}{\sqrt{4\pi D (t-t')}}
\end{equation}
The equation is also solved by the short time transition probability of the unprimed arguments $a(r_{ij},t)$ drift coefficient . The two-point short time transition probability can be set to a one-point probability  $p(r_{ij},t)=\frac{e^{-\frac{(r_{ij} - a'(r_{ij},t))}{2Dt}}}{\sqrt{4\pi D t}}$ and this is compared to the macroscopic maximum entropy information theory least biased distribution derived PDF....the diffusion coefficients being different in the square root of the potential, the powers of the $\mu, \nu$ as due to the partial differentiation of the force term, and the Lagrange multiplier of the potential term , which now we see can be obtained by inspection from this approach.

\pagebreak
 \section{Computation and numerical simulation.}
The numerics of computation for simulating chemical catalyzed glycol systems are simplified by these derived exact solutions. The distributions are then utilized with Monte Carlo utilizing PDFs  or stochastic trajectories utilizing the stochastic differential equations and numerical computed results obtained. As the distributions are exact solutions for approximated dynamics of interactions of the Lennard-Jones type, results can be calculated directly and ensemble simulations performed efficiently. Also, the NIST challenge, of obtaining a theoretical description for dissolved glycol in water and subsequent thermodynamic and phase properties can be either described analytically or simulated by numerics.

The parameters that are to be obtained either from fitting to experiment data  or by alternative means of energetics theory are the powers of the potential energy, $\mu, \nu$ of the Lennard-Jones potential $LJ(r_{ij})=\frac{a}{r^{\mu} _{ij}}- \frac{b}{ r^{\nu} _{ij} }$ and the Lagrange multiplier that sets the units of proportionality for the potential energy and coordinates....this last is a constant presumably as the time dependence is included with the variance or second moment's Lagrange multiplier. this can also be determined from experiment or energetics of thermodynamical relations.

\begin{figure}
\includegraphics[width=150mm] {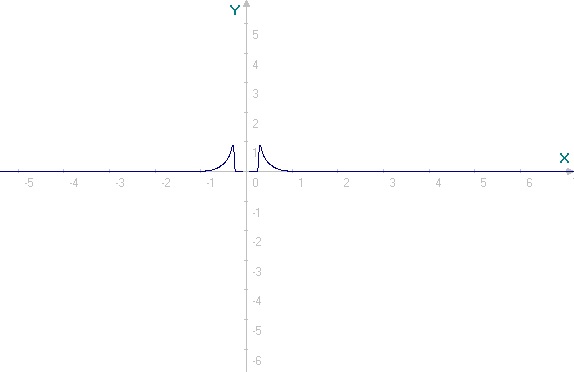} \label{fig1}
Fig. 1. p(x,t) vs. x , with y=p(x,t) and x= $r_{ij}$, the diffusion coefficient $D=0.001$, and time $t=1$ and $\mu=6, \nu=12$, $a,b=1$ for a Lennard-Jones potential estimation of the distribution behavior... note that as $x -> 0$ in arbitrary units the distribution decays rapidly to zero as expected, and as x becomes larger the probability increases, reaches symmetrically a maximum , and with increasing separation the probability of any two molecules interacting again decays to zero, though slowly with separation distance.
\end{figure}

\section{Conclusion}
In this article we have derived a general $\mu,\nu$ form of the $6,12$ Lennard-Jones -like fluid theory of Glycol water fluid statistical dynamics from an information theory and equivalently maximum entropy approach. This derivation is of the theoretical approach of modeling Hydrogen bonds interactions by Lennard-Jones and Lennard-Jones -like potentials. We have derived analogously to the fully quantum mechanical theory of phase transitions of fluids, vapors and solids and the matrix theory of interaction-formation of molecules from atomic components a matrix of probability distributions and therefore of the observables of the macroscopic dynamics of ether or ethylene and water molecules formed from Hydrogen, Oxygen, Carbon etc. that are in a particular fluid interaction phase at any given time.

   This is derived from first principles of atoms forming by interactions the molecules of the water, atoms forming the dissolved adducts of the glycol...the theory includes vapor-fluid transitions or evaporation.  We then solve analytically a particular distribution function of the interacting Hydrogen bonds' Lennard-Jones -like potential and obtain analytic forms that are normalized. This analytic form is due to the radial dependence of the Lennard-Jones potential and we compare the macroscopic derivation to the stochastic derivation by which the equivalent PDF with drift coefficient is obtained. We then discuss future research directions of computational simulations performed on fluids which the dissolved glycol water fluid is expected to have reduced freezing point, and which has been reported elsewhere and as cited here \cite{steuten1} which we will report in detail on for materials of water and glycol for simplicity.

\pagebreak
 *Fredrick Michael. Michael Research R$\&$D. fmicha3@uic.edu. fnmfnm2@yahoo.com. 773-641-0894.

\end{document}